\documentclass[twocolumn,preprintnumbers,amsmath,amssymb,prl]{revtex4}

\usepackage{graphicx}
\usepackage{dcolumn}
\usepackage{bm}
\usepackage{amssymb}
\usepackage{epstopdf}
\usepackage{color}

\begin{document}

\title{Fano resonances in optical spectra of semiconductor quantum wells dressed by circularly polarized light}
\author{O. V. Kibis$^1$}\email{Oleg.Kibis(c)nstu.ru}
\author{S. A. Kolodny$^1$}
\author{I. V. Iorsh$^{1,2}$}

\affiliation{$^1$Department of Applied and Theoretical Physics,
Novosibirsk State Technical University, Karl Marx Avenue 20,
Novosibirsk 630073, Russia}
\affiliation{$^2$Department of Physics
and Engineering, ITMO University, Saint Petersburg 197101, Russia}

\begin{abstract}
Optical properties of semiconductor quantum wells irradiated by a strong circularly polarized electromagnetic field are studied theoretically. Since the field can induce the composite electron-light states bound at repulsive scatterers, it drastically modifies all optical characteristics of the system. Particularly, it is demonstrated that the quantum interference of the direct interband optical transitions and the transitions through the light-induced intermediate states leads to the Fano resonances in the optical spectra, which can be detected in the state-of-the-art measurements.
\end{abstract}

\maketitle

The optical methods to control physical properties of solids
attract enormous attention of both optical and condensed-matter
scientific communities. Particularly, the engineering of various
quantum systems by an off-resonant electromagnetic field (Floquet
engineering) became the established research area which resulted
in many fundamental effects (see, e.g.,
Refs.~\cite{Goldman_2014,Bukov_2015,Basov_2017,Vogl_2019,
Vogl_2020}). Since the frequency of the off-resonant field lies
far from the optical absorption range, the field cannot be
absorbed and only ``dresses'' electrons (dressing field), changing
their physical characteristics. The effects induced by the
dressing field were actively studied during last years both
experimentally and theoretically in various nanostructures,
including quantum
rings~\cite{Kozin_2018,Kozin_2018_1},
quantum wells~\cite{Dini_2016,Kyriienko_2017},
topological
insulators~\cite{Lindner_2011,Morimoto_2016},
graphene and related 2D
materials~\cite{Oka_2009,Perez_2014,Kibis_2017,Iorsh_2017,Iurov_2020,Cavalleri_2020},
etc. Recently, we showed that the irradiation of 2D electron
systems by intensive circularly polarized light (dressing field)
induces the composite electron-light states, which are bound at repulsive scatterers~\cite{Kibis_2019,Kibis_2020}. Physical properties of the states are exceptional since repulsive potentials, normally, cannot confine electrons. As a consequence, they result in various novel quantum phenomena, including the
light-induced electron pairing~\cite{Kibis_2019} and the
Breit-Wigner resonance in electron transport~\cite{Kibis_2020}.
In the present Letter, we studied theoretically the
optical effects originated from
these light-induced electron states and found that they drastically modify the optical spectra of semiconductor quantum wells (QWs) by the Fano resonances.

\begin{figure}[!h]
\centering\includegraphics[width=7.7 cm]{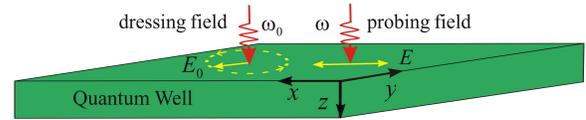}
\caption{Sketch of the system under consideration: Semiconductor
quantum well (QW) irradiated by a two-mode electromagnetic wave
consisting of a strong circularly polarized electromagnetic wave
with the electric field amplitude $E_0$ and the frequency
$\omega_0$ (dressing field) which induces the quasi-stationary
electron states bound at repulsive scatterers and a weak linearly
polarized electromagnetic wave with the electric field amplitude
$E$ and the frequency $\omega$ (probing field) which serves to
detect these states.}\label{Fig.1}
\end{figure}
Let us consider a semiconductor QW confining
electrons in the $x,y$ plane within the area $S$, where the
conduction band is empty, while the valence band is filled by
electrons. The QW is irradiated by the two-mode electromagnetic
wave (EMW) propagating along $z$ axis perpendicularly to the QW
plane (see Fig.~1). The first mode is a strong off-resonant
circularly polarized EMW (dressing field) with the vector
potential $\mathbf{A}(t)=(A_x,A_y)=[cE_0/\omega_0](\sin\omega_0
t,\,\cos\omega_0 t)$, where $E_{0}$ is the electric field
amplitude, and $\omega_0$ is the frequency which lies far from
characteristic resonant frequencies of the QW (particularly,
$\hbar\omega_0<\varepsilon_g$, where $\varepsilon_g$ is the band
gap of the QW). Since the dressing field is off-resonant, it
cannot be absorbed by the QW and only mixes electron states within
the conduction band. It should be noted also that the photon
energy of the off-resonant dressing field, $\hbar\omega_0$, should
be less than the inter-subband energy interval in QW to avoid the
inter-subband absorption of the dressing field. The second mode is
a weak resonant EMW (probing field) with the electric field
amplitude $E$ and the frequency $\omega$, which is linearly
polarized along the $x$ axis ($\hbar\omega>\varepsilon_g$). The
resonant probing field induces electron transitions between the
conduction and valence bands of the QW and serves to detect the
features of optical spectra originated from the dressing field.

In the QW dressed by the field $\mathbf{A}(t)$, the behavior of an electron near a scatterer with the repulsive potential $U(\mathbf{r})$ is described by the time-dependent Hamiltonian $\hat{\cal H}(t)=(\hat{\mathbf p}-e\mathbf{A}(t)/c)^2/2m_e+U(\mathbf{r})$, where $\hat{\mathbf p}$ is the plane momentum operator, and $m_e$ is the effective electron mass in the QW. If the field frequency $\omega_0$ is high enough, this time-dependent Hamiltonian can be reduced to the effective stationary Hamiltonian, $\hat{\cal H}_0=\hat{\mathbf p}^2/2m_e+U_0(\mathbf{r})$, where $U_0(\mathbf{r})=({1}/{2\pi})\int_{-\pi}^{\pi}U\big(\mathbf{r}-\mathbf{r}_0(t)\big)\,d(\omega_0
t)$ is the repulsive potential renormalized by the dressing field (dressed potential), $\mathbf{r}_0(t)=(-r_0\cos\omega_0 t,\,r_0\sin\omega_0 t)$ is the radius-vector describing the classical circular trajectory of a free electron in the circularly polarized field, and $r_0={|e|E_0}/{m_e\omega^2_0}$ is the radius of the trajectory~\cite{Kibis_2019,Kibis_2020}. In the case of the short-range scatterers conventionally modelled by the repulsive delta potential, $U(\mathbf{r})=u_0\delta(\mathbf{r})$, the corresponding dressed potential reads~\cite{Kibis_2020}
\begin{equation}\label{U0}
U_0(\mathbf{r})=\frac{u_0\,\delta({r}-{r}_0)}{2\pi r_0}.
\end{equation}
Thus, the circularly polarized dressing field turns the repulsive
delta potential into the
delta potential barrier of ring shape (\ref{U0}) pictured in
Fig.2a, which defines dynamics of an electron dressed by the field. As a consequence, the bound electron states which are
confined inside the area fenced by the ring-shape barrier
($0<r<r_0$) appear. Certainly,
the bound electron states are quasi-stationary since they can
decay via the tunnel transition through the potential barrier into
the continuum of conduction electrons. As a consequence, the
energy broadening of the bound states appears. Assuming the repulsive delta potential to be strong ($\alpha=2\hbar^2/m_eu_0\ll1$), the solution of the Schr\"odinger problem with the stationary potential (\ref{U0}) results in
the energy spectrum of the bound quasi-stationary states,
\begin{equation}\label{Elm}
\varepsilon_{nm}=\frac{\hbar^2\xi^2_{nm}}{2m_er^2_0}+{\cal
O}\left(\alpha\right),
\end{equation}
their energy broadening,
\begin{equation}\label{Glm}
{\Gamma}_{nm}=\frac{4\varepsilon_{nm}\alpha^2}{N^3_m(\xi_{nm})[J_{m+1}(\xi_{nm})-J_{m-1}(\xi_{nm})]}+{\cal
O}\left(\alpha^3\right),
\end{equation}
and their wave functions,
\begin{eqnarray}\label{Flm}
\psi_{nm}&=&\frac{e^{i3\pi/4}e^{im\varphi}}{\sqrt{\pi}r_0J_{m+1}(\xi_{nm})}\left\{
\begin{array}{rl}
J_m\left(\frac{\xi_{nm}r}{r_0}\right), &0<r\le r_0\\
0, &r\ge r_0
\end{array}\right.\nonumber\\
&+&{\cal O}\left(\alpha\right),
\end{eqnarray}
where $J_m(\xi)$ and $N_m(\xi)$ are the Bessel functions of the
first and second kind, respectively, $\xi_{nm}$ is the $n$th zero
of the Bessel function $J_m(\xi)$, $n=1,2,3,...$ is the principal
quantum number which numerates zeros of the Bessel function
$J_m(\xi)$, and $m=0,\pm1,\pm2,...$ is the angular momentum~\cite{Kibis_2020}. The
ground bound quasi-stationary state with the energy
$\varepsilon_{10}$ is pictured schematically in Fig.~2a, where the
tunnel transition from this state to the continuum of free
conduction electrons is marked by the arrow.
\begin{figure}[!h]
\centering\includegraphics[width=7.7 cm]{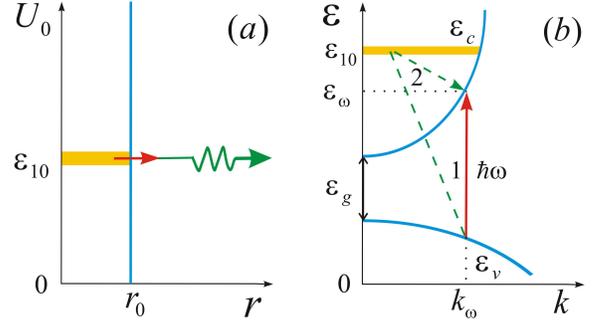}
\caption{Scheme of electron transitions: (a) tunnel transition
(the red solid arrow) from the ground quasi-stationary state with
the energy $\varepsilon_{10}$ (the horizontal yellow strip) to the
state of free conduction electron (the green wave  arrow) through
the ring-shape delta potential barrier
$U_0=u_0\delta({r}-{r}_0)/2\pi r_0$ (the vertical blue line); (b)
optical transitions from the valence band $\varepsilon_v$ to the
conduction band $\varepsilon_c$ induced by the probing field with
the frequency $\omega$: the direct transition (the solid arrow
$1$) and the transition through the quasi-stationary state with
the energy $\varepsilon_{10}$ (the dashed arrow
$2$).}\label{Fig.2}
\end{figure}

For applicability of Eqs.~(\ref{Elm})--(\ref{Flm}) to the considered problem, the two conditions are assumed to be satisfied:
Firstly, $\omega_0\tau_e\gg1$,
where $\tau_e$ is the mean free time of conduction electrons;
secondly, the field frequency, $\omega_0$, lies far from the resonant frequencies corresponding to the optical transitions between the energy levels (\ref{Elm}). Physically, the first condition allows to neglect the scattering
processes which can destroy the bound states~(\ref{Elm})--(\ref{Flm}), whereas the second
condition allows to neglect the effect of the oscillating terms of
the total Hamiltonian $\hat{\cal H}(t)$, which were omitted under the transformation of this time-dependent Hamiltonian to the stationary form $\hat{\cal H}_0$ (see Ref.~\cite{Kibis_2020} for more details).

Let us consider optical spectra of the QW near the
optical-absorption edge, where the probing field induces the
optical electron transitions between the first valence subband and
the first conduction subband (see Fig.~2b). Then the energy
spectrum of conduction (valence) electrons is
$\varepsilon_{c(v)}(\mathbf{k})=\varepsilon_{c(v)0}+\varepsilon_{c(v){k}}$,
where $\varepsilon_{c\,0,v0}$ are the energies of the subband
edges, $\varepsilon_{c{k}}=\hbar^2 k^2/2m_e$ and
$\varepsilon_{v{k}}=-\hbar^2 k^2/2m_h$ are the energies of
electrons within the conduction and valence subbands,
respectively, $\mathbf{k}=(k_x,k_y)$ is the electron wave vector
and $m_{e(h)}$ are the effective masses of electrons (holes) in
the subbands. In the following, the basic electron states
corresponding to these energies will be denoted as
$|\mathbf{k}_{c(v)}\rangle=|\sqrt{S}\,e^{i\mathbf{kr}}\phi_{c(v)}\rangle$,
where $\phi_{c(v)}$ are the wave functions corresponding to the
subband edges (they include both the Bloch functions of the
semiconductor material and the subband wave function arisen from
the size quantization in the QW). The quasi-stationary electron
states bound at different scatterers will be denoted as
$|s^{(j)}_{nm}\rangle=|\psi^{(j)}_{nm}\phi_{c}\rangle$, where the
index $j=1,2,...N$ numerates scatterers located in different
places of the $x,y$ plane and $N$ is the total number of
scatterers in the QW. The considered two-band electron system interacting with the
two-mode electromagnetic field can be described by the effective
Hamiltonian
$\hat{\cal H}_\mathrm{eff}=\hat{\cal H}_e+\hat{\cal
H}_T+E\cos\omega t\,\hat{\cal H}_D$,
where
\begin{align}\label{He}
&\hat{\cal
H}_e=\sum_{j=1}^N\sum_{n,m}|s^{(j)}_{nm}\rangle\varepsilon_{nm}\langle
s^{(j)}_{nm}|+\sum_{\mathbf{k}_c}|\mathbf{k}_c\rangle\varepsilon_{ck}\langle
\mathbf{k}_c|\nonumber\\
&+\sum_{\mathbf{k}_v}|\mathbf{k}_v\rangle\varepsilon_{vk}\langle
\mathbf{k}_v|\nonumber
\end{align}
is the electron Hamiltonian describing energies of the basic
electron states,
\begin{equation}\label{HT}
\hat{\cal
H}_T=\sum_{j=1}^N\sum_{n,m}\sum_{\mathbf{k}_c}|\mathbf{k}_c\rangle\langle\mathbf{k}_c|\hat{\cal
H}_T|s^{(j)}_{nm}\rangle \langle s^{(j)}_{nm}|+\mathrm{H.c.}\nonumber
\end{equation}
is the tunnel Hamiltonian describing the tunnel transitions from
the quasi-stationary bound states to the states of conduction
electrons through the ring-shape potential barrier (see Fig.~2a),
and the Hamiltonian of the dipole interaction between electron
states in the valence band and the conduction band induced by the
probing field is
\begin{eqnarray}\label{HD}
\hat{\cal
H}_D&=&\sum_{\mathbf{k}_v}\Big[\,\sum_{\mathbf{k}_c}|\mathbf{k}_c\rangle\langle
\mathbf{k}_c|ex|\mathbf{k}_v\rangle\langle\mathbf{k}_v|\nonumber\\
&+&\sum_{j=1}^N\sum_{n,m}|s^{(j)}_{nm}\rangle\langle
s^{(j)}_{nm}|ex|\mathbf{k}_v\rangle\langle\mathbf{k}_v|+\mathrm{H.c.}\,\Big]\nonumber
\end{eqnarray}

Assuming the tunneling between the quasi-stationary electron states and the states of conduction
electrons to be weak and the probing field amplitude, $E$, to be
small, the last two terms of the Hamiltonian $\hat{\cal H}_\mathrm{eff}$ can be
considered as a perturbation. Then the probability of optical
interband electron transition from the state
$|\mathbf{k}^\prime_v\rangle$ to the state $|\mathbf{k}_c\rangle$
per unit time reads~\cite{Landau_3}
\begin{align}\label{Wkk}
&w_{\mathbf{k}_c\mathbf{k}^\prime_v}=\frac{\pi E^2}{2\hbar}\,\delta(\varepsilon_{c{k}}+\varepsilon_{v{k}^\prime}+\varepsilon_g-\hbar\omega)\nonumber\\
&\times\left|\langle\mathbf{k}_c|\hat{\cal
H}_D|\mathbf{k}^\prime_v\rangle+\sum_{j=1}^{N}\sum_{n,m}\frac{\langle\mathbf{k}_c|\hat{\cal
H}_T|s^{(j)}_{nm}\rangle\langle s^{(j)}_{nm}|\hat{\cal
H}_D|\mathbf{k}^\prime_v\rangle}{\varepsilon_{c0}+\varepsilon_{ck}-\varepsilon_{nm}+i\Gamma_{nm}/2}\right|^2,
\end{align}
where the energy broadening $\Gamma_{nm}$ can be written as
\begin{eqnarray}\label{Gamma0}
\Gamma_{nm}=\frac{Sm_e}{\hbar^2}\left|\langle\mathbf{k}_{nm}|\hat{\cal
H}_T|s_{nm}\rangle\right|^2,
\end{eqnarray}
$|\mathbf{k}_{nm}\rangle$ is the state of conduction electron with
the bound state energy $\varepsilon_{nm}=\hbar^2k^2_{nm}/2m_e$,
and $|s_{nm}\rangle$ is the quasi-stationary electron state with
the energy $\varepsilon_{nm}$ confined at a scatterer positioned
in the zero point of the coordinate system. In what follows, the
matrix elements in Eq.~(\ref{Wkk}) will be rewritten as
$\langle\mathbf{k}_c|\hat{\cal H}_T|s^{(j)}_{nm}\rangle\langle
s^{(j)}_{nm}|\hat{\cal
H}_D|\mathbf{k}^\prime_v\rangle=e^{i(\mathbf{k}^\prime_v-\mathbf{k}_c)\mathbf{R_j}}\langle\mathbf{k}_c|\hat{\cal
H}_T|s_{nm}\rangle\langle s_{nm}|\hat{\cal
H}_D|\mathbf{k}^\prime_v\rangle$, where $\mathbf{R}_j$ is the
radius vector of $j$th scatterer position.

The first term under the modulus in the probability (\ref{Wkk})
describes the usual direct interband transition (see the vertical
solid arrow $1$ in Fig.~2b), whereas the second term corresponds
to the transitions through the intermediate quasi-stationary
states (see the dashed arrow $2$ in Fig.~2b, which marks such a
transition through the ground quasi-stationary state
$\varepsilon_{10}$). It follows from the energy conservation law
that the transitions through the quasi-stationary states are
possible only within the narrow energy range of conduction
electrons,
$\varepsilon_{ck}\approx\varepsilon_{nm}\pm\Gamma_{nm}/2$. If the
broadening $\Gamma_{nm}$ is small, the matrix elements
$\langle\mathbf{k}_c|\hat{\cal H}_T|s_{nm}\rangle$ in
Eq.~(\ref{Wkk}) varies little around the energy
$\varepsilon_{ck}=\varepsilon_{nm}$ for which the probability of
the transition is not negligible. Therefore, one can replace the
tunnel matrix elements $\langle\mathbf{k}_c|\hat{\cal
H}_T|s_{nm}\rangle$ in Eq.~(\ref{Wkk}) with the resonant matrix
elements, $\langle\mathbf{k}_{nm}|\hat{\cal H}_T|s_{nm}\rangle$.

Taking into account Eq.~(\ref{Gamma0}) and the solutions of the
Schr\"odinger problem (\ref{Elm})--(\ref{Flm}), the resonant
matrix element of the conventional tunnel
Hamiltonian~\cite{Bardeen_1961} can be written as
$\langle\mathbf{k}_n|\hat{\cal H}_T|s_{nm}\rangle=-\hbar
\sqrt{\Gamma_{nm}/Sm_e}$. Then the probability (\ref{Wkk}) can be
rewritten as
\begin{align}\label{WF}
&w_{\mathbf{k}_c\mathbf{k}^\prime_v}=\,\delta(\varepsilon_{c{k}}+\varepsilon_{v{k}^\prime}+\varepsilon_g-\hbar\omega)\times\Bigg|4\pi^2\delta(\mathbf{k}-\mathbf{k}^\prime)\,-\nonumber\\
&\left.\frac{\hbar}{\sqrt{m_e}}\sum_{j=1}^N\sum_{n,m}
\frac{{\Gamma^{1/2}_{nm}}\Phi_{nm}(\mathbf{k}^\prime)e^{i(\mathbf{k}^\prime-\mathbf{k})\mathbf{R_j}}}
{\varepsilon_{c0}+\varepsilon_{ck}-\varepsilon_{nm}+i\Gamma_{nm}/2}\right|^2\frac{\pi|D_{cv}|^2E^2}{2\hbar
S^2},
\end{align}
where $\Phi_{nm}(\mathbf{k})=\langle
\psi_{nm}|e^{i\mathbf{kr}}\rangle$ is the Fourier transform of the
bound state wave functions (\ref{Flm}), $D_{cv}=(\hbar
em_0/iS\varepsilon_g)\langle\phi_c|\hat{p}_x|\phi_v\rangle$ is the
interband matrix element of electric dipole moment, and $m_0$ is
the electron mass in vacuum. Next, we have to average the
probability (\ref{WF}) over coordinates of all $N$ scatterers.
Assuming the scatterers to be randomly arranged in the QW plane,
the averaging procedure is defined by the operator
$\hat{A}=(1/S)^N\prod_{j=1}^N\left[\int_Sd^2\mathbf{R}_j\right]$.
Then the intensity of absorption of the probing field,
$I=(\hbar\omega/S)\sum_{\mathbf{k}_c,\mathbf{k}^\prime_v}
\hat{A}w_{\mathbf{k}_c\mathbf{k}^\prime_v}$, reads
\begin{align}\label{Ia}
&I=\int_{\mathbf{k}^\prime_v} d^2\mathbf{k}^\prime\int_{\mathbf{k}_c} d^2\mathbf{k}\Bigg(\frac{\hbar^2n_s[1+4\pi^2n_s\delta(\mathbf{k}-\mathbf{k}^\prime)]}{m_e}\nonumber\\
&\times
\left|\sum_{n,m}\frac{{\Gamma^{1/2}_{nm}}\Phi_{nm}(\mathbf{k}^\prime)}
{\varepsilon_{c0}+\varepsilon_{ck}-\varepsilon_{nm}+i\Gamma_{nm}/2}\right|^2+4\pi^2
\delta(\mathbf{k}-\mathbf{k}^\prime)\nonumber\\
&-\left.\frac{8n_s\pi^2\hbar}{\sqrt{m_e}}\mathrm{Re}\left[\sum_{n,m}\frac{\Gamma^{1/2}_{nm}\Phi_{nm}(\mathbf{k}^\prime)}
{\varepsilon_{c0}+\varepsilon_{ck}-\varepsilon_{nm}+i\Gamma_{nm}/2}\right]\right.\nonumber\\
&\times\delta(\mathbf{k}-\mathbf{k}^\prime)\Bigg)\left(\frac{\omega|D_{cv}|^2E^2}{16\pi^3}\right)\delta(\varepsilon_{ck}+\varepsilon_{vk^\prime}+\varepsilon_g-\hbar\omega),
\end{align}
where $n_s=N/S$ is the density of quasi-stationary states (density
of scatterers). Restricting the consideration by the ground
quasi-stationary state ($n=1,\,m=0$) which defines the low-energy
optical properties, the absorption intensity (\ref{Ia}) can be
rewritten as
\begin{align}\label{IF}
&I=\left(\frac{\omega m_e|D_{cv}|^2E^2}{2\hbar^2}\right)
\Bigg[1+\frac{\hbar^2n^2_s{\Gamma_{10}}[|\Phi_{10}(k_\omega)|^2+n_s\bar{\Phi}_{10}]}
{m_e[(\varepsilon_{\omega}-\varepsilon_{10})^2+(\Gamma_{10}/2)^2]}\nonumber\\
&-\left.\frac{{2}n_s\hbar\Gamma^{1/2}_{10}\Phi_{10}(k_\omega)(\varepsilon_{\omega}-\varepsilon_{10})}
{\sqrt{m_e}\,[(\varepsilon_{\omega}-\varepsilon_{10})^2+(\Gamma_{10}/2)^2]}\right],
\end{align}
where
$k_\omega=\sqrt{2(\hbar\omega-\varepsilon_g)m_em_h/(m_e+m_h)}/\hbar$
is the resonant electron wave vector corresponding to the direct
interband optical transition (see Fig.~2b),
$\varepsilon_\omega=\varepsilon_{c0}+\hbar^2k^2_\omega/2m_e$ is
the resonant energy in the conduction band (see Fig.~2b), and
$\bar{\Phi}_{10}=\int_0^{\infty}|\Phi_{10}(k)|^2kdk$. Substituting
Eqs.~(\ref{Elm})--(\ref{Flm}) into Eq.~(\ref{IF}),
one can find the dependence of the absorption spectrum of QW on
the dressing field intensity, $I_0=cE_0^2/4\pi$, which is plotted
in Fig.~3 for GaAs-based quantum well (the effective mass of
electrons is $m_e\approx0.0067\,m_0$ and the effective mass of
holes in the first subband~\cite{Dyakonov_1982} is
$m_h\approx0.11\,m_0$).
\begin{figure}[!h]
\centering\includegraphics[width=7.7 cm]{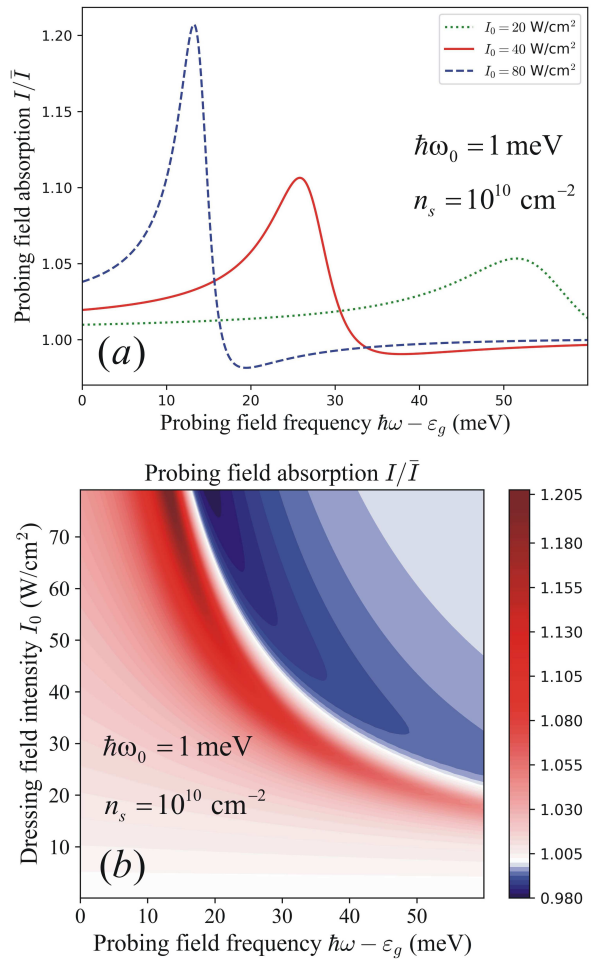}
\caption{Optical absorption spectra of the probing field in a
GaAs-based quantum well with the scatterer density
$n_s=10^{10}$~cm$^{-2}$ and the energy broadening
$\Gamma_{10}=0.1\varepsilon_{10}$ for the dressing field photon
energy $\hbar\omega_0=1$~meV and different dressing field
intensities, $I_0$, in the units of optical absorption without the
dressing field, $\bar{I}$.}\label{Fig.3}
\end{figure}

The first term in the square brackets of Eq.~(\ref{IF}) arises
from the direct optical transition (see the solid arrow $1$ in
Fig.~2b), which does not depend on the probing field frequency,
$\omega$, since the density of electron states near edges of 2D
subbands does not depend on the electron energy. Just this term
describes the intensity of optical absorption in the absence of
the dressing field, $\bar{I}={\omega
m_e|D_{cv}|^2E^2}/{2\hbar^2}$. The second term there arises from
optical absorption through the quasi-stationary state
$\varepsilon_{10}$ (see the dashed arrow $2$ in Fig.~2b) and is
described by the Lorentzian centered at the resonant energy
$\varepsilon_\omega=\varepsilon_{10}$, whereas the third term
arises from the quantum interference of the absorption ways $1$
and $2$ in Fig.~2b and depends on the broadening of
quasi-stationary state $\Gamma_{10}$. Since the interference term
changes its sign at the resonant energy
$\varepsilon_\omega=\varepsilon_{10}$, we arrive at the
asymmetrical structure of the total absorption spectrum plotted in
Fig.~3, which is typical for the Fano resonances~\cite{Fano_1961}.
In the present plots, we restricted the consideration by the
resonance arisen from the ground quasi-stationary state
$\varepsilon_{10}$. Certainly, analogous Fano resonances will
appear from other quasi-stationary states (overlying in energy) in
the high-frequency area of the absorption spectrum.

Within the developed theoretical model, we focused exclusively on optical features originated from the quasi-stationary electron states induced by the dressing field. Certainly, experimentally observable optical spectra of semiconductor QWs contain a lot of details arisen from other phenomena (particularly, excitonic effects). However, the Fano resonance is characteristic only for the optical absorption through the quasi-stationary states and, therefore, the discussed effect can be easily detected in measurements. It should be noted also that the dressing field can induce the dynamical Franz-Keldysh (FK) effect, which modifies the joint density of states, resulting in the formation of the side bands in the vicinity of the fundamental absorption edge~\cite{Jauho_1996}. However, the FK modifications of spectra take place only in the energy region $\sim\hbar\omega_0$ near the band edge, whereas the Fano resonance appears in the area of resonant electron energies $\sim10\hbar\omega_0$  (see Fig. 3a). Moreover, the amplitude of the dynamical FK effect is very small for the considered field intensities of tens W/cm$^2$. Thus, manifestations of the dynamical FK effect~\cite{Jauho_1996} in our system are qualitatively different from the Fano effect under consideration.

Summarizing the aforesaid, one can conclude that the quasi-stationary electron states induced by a circularly polarized dressing field  manifest themselves in optical spectra of the
QWs as the Fano resonances. It is demonstrated that the resonance
peaks are positioned at energies of the states, whereas the Fano
asymmetry of the peaks depends on the energy broadening of the
quasi-stationary states. Therefore, the present theory allows to
use optical measurements as an experimental method to detect
energy structure of the light-induced electron states.

{\bf Funding.}
The reported study was funded by the Russian Science Foundation
(project 20-12-00001).

{\bf Acknowledgment.} Portions of this work were presented at the 5th International Conference on Metamaterials and Nanophotonics METANANO in 2020, paper number 020052.

{\bf Disclosures.}
The authors declare no conflicts of interest.

\end{document}